\documentclass[a4paper,fleqn,usenatbib,twocolumn]{mnras}
\usepackage{pdfpages}
\usepackage[T1]{fontenc}
\usepackage{ae,aecompl}
\usepackage{graphicx}
\usepackage{amsmath}
\usepackage{amssymb}
\usepackage{epstopdf}
\usepackage{subfig}
\usepackage{natbib}

\defcitealias{Hamish1}{Paper I}

\title[Dark matter and pulsars: II. Small-scale cosmology]{Investigating dark matter substructure with pulsar timing\\ II. Improved limits on small-scale cosmology}

\author[Hamish A. Clark et al.]{
Hamish A. Clark,$^{1}$\thanks{Email: hamish.clark@sydney.edu.au (HAC)}
Geraint F. Lewis,$^{1}$
Pat Scott$^{2}$
\\
$^{1}$Sydney Institute for Astronomy, School of Physics A28, The University of Sydney, NSW 2006, Australia\\
$^{2}$Department of Physics, Imperial College London, Blackett Laboratory, Prince Consort Road, London SW7 2AZ, UK
}

\date{Accepted XXX. Received YYY; in original form ZZZ}

\pubyear{2015}

\begin{document}
\label{firstpage}
\pagerange{\pageref{firstpage}--\pageref{lastpage}}
\maketitle

\begin{abstract}
Ultracompact Minihalos (UCMHs) have been proposed as a type of dark matter sub-structure seeded by large-amplitude primordial perturbations and topological defects.  UCMHs are expected to survive to the present era, allowing constraints to be placed on their cosmic abundance using observations within our own Galaxy. Constraints on their number density can be linked to conditions in the early universe that impact structure formation, such as increased primordial power on small scales, generic weak non-Gaussianity, and the presence of cosmic strings. We use new constraints on the abundance of UCMHs from pulsar timing to place generalised limits on the parameters of each of these cosmological scenarios. At some scales, the limits are the strongest to date, exceeding those from dark matter annihilation. Our new limits have the added advantage of being independent of the particle nature of dark matter, as they are based only on gravitational effects.
\end{abstract}




\section{Introduction}
Rare objects seeded by large density contrasts at early times are an effective tool for probing the early universe. By determining the present-day abundance of such objects, we can investigate the processes leading to their production. Examining the formation of different objects allows us to test different physical processes, scales, and epochs. This probe of the early universe is unique in its approach, allowing for constraints to be placed on cosmological parameters at scales far smaller than would otherwise be accessible.

Primordial black holes (PBHs) are an extreme example of such objects, and have long been used as a probe of the small-scale universe. Should a primordial fluctuation exceed a large threshold value at horizon entry ($\delta \gtrsim 0.3$), the force of gravitation will overcome that of pressure and the region will collapse, forming a black hole \citep{Carr-Hawking, Carr}. Constraints on the production of primordial black holes have been obtained from a multitude of methods, and have been used to weakly constrain curvature perturbations \citep{Josan, Carr-Kohri, Alabidi}, non-Gaussianity \citep{Young13, Shandera, Young15a, Young15b} and inflation \citep{Bringmann02,PeirisPBH} over a relatively large range of scales.

In cases where density fluctuations are larger than $\delta \sim 10^{-3}$ but too small to form a PBH, the dark matter contained in the perturbation is expected to collapse so quickly that an ultracompact minihalo (UCMH) would form \citep{Berezinsky2003, Ricotti2009, SS09}.  UCMHs are distinguished from regular dark matter structure by their very early time of collapse, around the time of matter-radiation equality or even earlier \citep{Berezinsky12}.  Consequently, UCMHs have extremely steep density profiles, and are expected to persist to the present day, as they would not be easily tidally disrupted \citep{Berezinsky2006, Berezinsky2008, Bringmann}.  It has been shown that limits on the abundance of UCMHs can be mapped to corresponding limits on processes that are expected to increase their production: increased primordial power on small scales \citep{JG10, Bringmann}, non-Gaussianity \citep{Shandera}, and the presence of cosmic strings in the early Universe \citep{Berezinsky2011, Anthonisen}.  UCMHs have also been studied extensively for their promise as sources of dark matter annihilation or decay \citep{SS09, Lacki10, Yang11c, Yang12, Yang11a, Yang13a, Yang13c, Yang13b, Zhang11, Zheng14}.

To date, the strongest limits on the UCMH abundance have come from non-detection of dark matter clumps in gamma rays by the $Fermi$ Large Area Telescope (LAT), relying on the assumption that dark matter can annihilate \citep{Bringmann}. In \citetalias{Hamish1} \citep{Hamish1}, we showed that a population of UCMHs also will produce a detectable effect on the period derivative of pulsars, due to their gravitational time delay.  By exploiting this effect, we showed that purely gravitational arguments place a strong limit on the fraction of dark matter within the Milky Way contained within UCMHs.  These limits are significantly stronger than the only previous gravitational limits \citep{Zackrisson12}, which were placed by assuming non-detection of small distortions in the images of macrolensed quasar jets. Although they do not cover as broad a mass range, for some masses the pulsar limits are even stronger than those from gamma-ray searches.

Here we apply our new pulsar limits on the UCMH abundance to produce updated, fully model-independent constraints on cosmological scenarios that could give rise to UCMHs. By calculating the expected UCMH abundance for a given primordial power and scale, in Section \ref{PPS} we give generalised constraints on the small-scale primordial power, as well as on simple power-law spectra. As the production of rare objects has been seen to be very sensitive to higher moments of the distribution of primordial fluctuations \citep{Bullock, LoVerde08, Shandera}, in Section \ref{NG} we place limits on the amount of generic non-Gaussianity allowed on small scales.  Cosmic strings -- topological defects from symmetry-breaking phase transitions in the early Universe \citep[see e.g.][]{Brandenberger94} -- have also been shown to act as seeds for formation of dark matter substructure \citep{Berezinsky2011, Anthonisen}.  In Section \ref{CS}, we apply our new limits on the UCMH abundance to constrain the cosmic string tension. In what follows, unless stated otherwise, we closely follow the methods of \citet{Bringmann}, \citet{Shandera} and \citet{Anthonisen} for the respective cosmological scenarios.  The code used by each has been implemented in v5.1.2 of \textsc{DarkSUSY} \citep{Darksusy}, providing routines to compute the abundance of UCMHs independent of the assumed model of dark matter.

\section{Constraints on Primordial Power}
\label{PPS}
Primordial fluctuations are thought to have given rise to the large-scale structure of the universe. These density perturbations acted as the seeds for small-scale structures, which gravitationally collapsed and merged to form a network of sheets, filaments, and voids (for an overview see \citealt{Mo}). These fluctuations are very well constrained on large scales by many observations \citep{Mcdonald, Reid,Chluba,WMAP2013, Planck2013, Planck2015}, indicating that their power appears to be nearly equal on all such larger scales. However, given that the imprint of fluctuations has yet to actually be observed on small scales, mechanisms that would increase (or decrease) power on small scales by some degree are not disallowed.

While the PBH abundance has been used as a probe of curvature perturbations on very small scales (as small as $k\sim10^{19}$ Mpc$^{-1}$), this constraint is quite weak in comparison to others -- by a factor of approximately 7 orders of magnitude. Similarly, the UCMH abundance has been used to constrain curvature perturbations far more strongly ($\mathcal{P}_\mathcal{R} \lesssim 10^{-7}$ in the range $10 \lesssim k \lesssim 10^7$ Mpc$^{-1}$), using gamma-ray searches with the $Fermi$-LAT \citep{Bringmann}. From the limits on the present-day UCMH number density within the Milky Way that we found in \citetalias{Hamish1}, here we go beyond the assumption of annihilating dark matter and provide concrete, model-independent limits on primordial power at small scales.

The present day mass $M_h^0$ of a UCMH is related to the co-moving radius $R$ of the initial overdense region at horizon entry by \citep{Bringmann}
\begin{equation}
M^0_h \approx 4 \times 10^{13} \left(\frac{R}{\textrm{Mpc}}\right)^3 M_\odot.
\label{UCMHmass}
\end{equation}

The fraction of dark matter expected to be contained in UCMHs of mass $M_h^0$ at redshift $z$ is defined as
\begin{equation}
f(z) = \beta (R) f_\chi \frac{z_{eq}+1}{z+1},
\end{equation}
where $\beta (R)$ is the probability that such a region will seed the formation of a UCMH, $z_{eq}$ is the redshift of matter-radiation equality, and $f_\chi \equiv \Omega_\chi/\Omega_m$ is the fraction of matter that is cold dark matter (CDM). Here accretion of dark matter from the cosmological background onto UCMHs is taken to continue up to $z \sim 10$, after which structure formation has evolved such that the majority of halos will be within gravitationally bound systems. The present day fraction in the Milky Way will then be $f_{MW}\approx 250.77 \beta (R)$.

Assuming that the primordial perturbations follow a Gaussian distribution, the probability of UCMH formation may be found as
\begin{equation}
\beta(R) \simeq \frac{\sigma_{\chi, \textrm{H}}(R)}{\sqrt{2\pi}\delta_\chi^{min}}\exp \left[ -\frac{{\delta_\chi^{min}}^2}{2\sigma_{\chi, \textrm{H}}^2(R)}\right],
\label{beta}
\end{equation}
where $\sigma_{\chi, \textrm{H}}^2$ is the dark matter mass variance at horizon entry and $\delta_\chi$ is the minimum density contrast required to produce a UCMH.  The minimum density contrast is a function of both wavenumber $k$, and the latest allowed redshift of UCMH collapse, $z_c$ (see Appendix A in \citealt{Bringmann}). At redshifts $z<z_c$, the background dark matter will collapse with sufficient angular momentum that the radial infall approximation required to produce the steep density profile that characterises a UCMH no longer applies. As there is not yet a concrete understanding of what this latest redshift of collapse is, we display results for both a rather conservative estimate of $z_c = 1000$ and the slightly more liberal $z_c = 200$.

\begin{figure}
\centering
\includegraphics[width=\columnwidth]{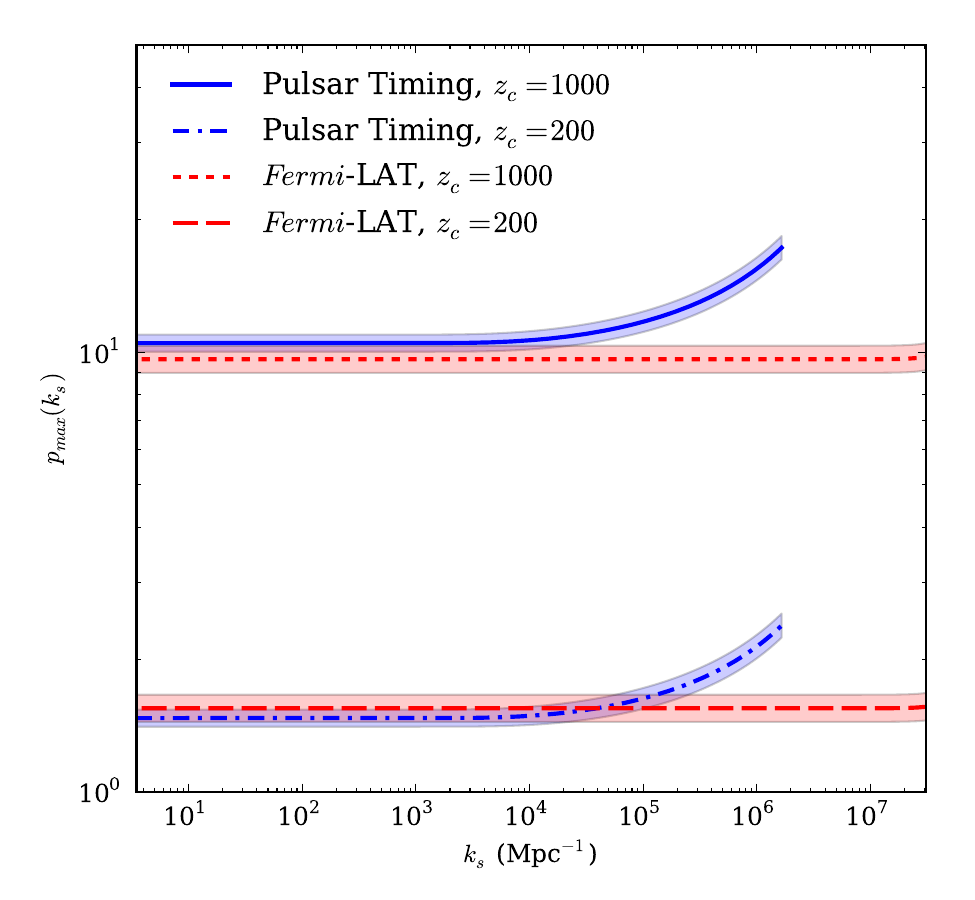}
\caption{Constraints on the step size for a stepped primordial spectrum of primordial fluctuations, as a function of step position. Limits correspond to an assumed spectral index of $n_s = 0.968$, as well as $1\sigma$ deviations allowed by observations from Planck \citep{Planck2015}. We show limits from both pulsar timing and gamma-ray searches, for redshifts of latest collapse $z_c = 1000$ and $z_c =200$.}
\label{steplims}
\end{figure}

Solving Eq.\ \ref{beta} by use of Brent's Method \citep{Brent}, in conjunction with the limits on $f_{MW}$ in \citetalias{Hamish1}, we find constraints on primordial mass variance, $\sigma_{\chi, \textrm{H}}^2$. To express the amplitude of a curvature perturbation $\mathcal{P}_\mathcal{R}$ in terms of the mass variance, a power spectrum model must be assumed. We follow the power spectrum normalisation described in Appendix B of \citet{Bringmann}, for three different models:
\begin{enumerate}
\item A `generalised' power spectrum, which assumes local scale invariance rather than the global invariance of the Harrison--Zel'Dovich model:
\begin{equation}
\mathcal{P}_\mathcal{R}(k) = \mathcal{P}_\mathcal{R}(k_R)\left(\frac{k}{k_R}\right)^{n_R(k_R)-1}.
\end{equation}
Here $n_R(k_R)$ is the \textit{local} slope of the power law at $k_R$, which we take to be $n_R =1$. It should be noted that the limits we derive are expected to change for $n_R \ne 1$, as it is not possible to relate mass variance and curvature entirely without model assumptions. This generalised power spectrum provides a normalisation of
\begin{equation}
\sigma_{\chi,\textrm{H}}^2(R) = 0.907 \mathcal{P}_\mathcal{R}(k),
\end{equation}
resulting in limits on primordial curvature $\mathcal{P}_\mathcal{R}$.

We show the resulting limits in Fig.\ \ref{curvlims}. For the case $z_c = 200$, these limits are of comparable strength to those obtained from large-scale observations ($\log_{10}\mathcal{P}_\mathcal{R} \lesssim -8.5$), but are extended to much larger $k$. We reiterate, however, that it it is not currently known if UCMH formation can continue up to this point, so the weaker limits ($z_c = 1000$) should be considered more robust.

\begin{figure}
\centering
\includegraphics[width=\columnwidth]{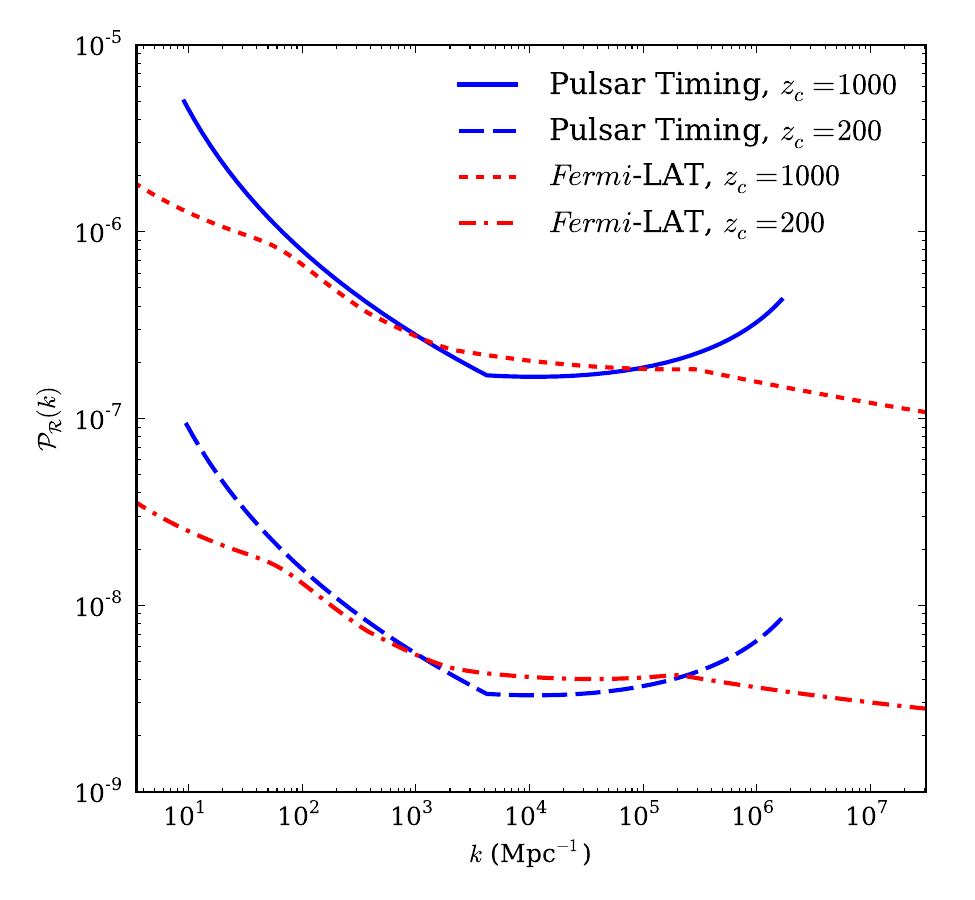}
\caption{Upper limits (at 95\% CL) on the amplitude of primordial curvature for a generalised power spectrum. We show those obtained from both gamma-ray searches and pulsar timing, and for two assumed latest allowed redshifts of UCMH collapse, $z_c = 200$ and $z_c = 1000$. The constraints obtained by gamma-ray searches are shown for an assumed dark matter mass of $m_\chi = 1$ TeV, which annihilates entirely into $b\bar{b}$ pairs with thermally-averaged cross section $\langle\sigma v\rangle = 3\times 10^{-26}$ $\textrm{cm}^3\textrm{s}^{-1}$.}
\label{curvlims}
\end{figure}

\item A scale-free spectrum with constant spectral index $n_s$:
\begin{equation}
\mathcal{P}(k) \propto k^{n_s - 1}.
\end{equation}

We again follow the method in \citet{Bringmann}, with our derived constraints on spectral index shown in Fig.\ \ref{nlims}. The appropriate limit to take from these constraints will be the lowest at any scale: $n_s \leq 1.24$ ($z_c = 1000$), and $n_s \leq 1.02$ ($z_c = 200$). Although neither of these constraints is as strong as the corresponding limit from gamma-ray searches ($n_s \leq 1.16$ for $z_c = 1000$, and $n_s \leq 1.00$ for $z_c = 200$), they apply without any assumptions about the specific particle nature of dark matter. Likewise, the limits on the scale-free spectral index from cosmic microwave background (CMB) observations are in agreement with those we find here (e.g.\ $n_s = 0.968 \pm 0.006$; \citealt{Planck2015}), but are markedly stronger.

\begin{figure}
\centering
\includegraphics[width=\columnwidth]{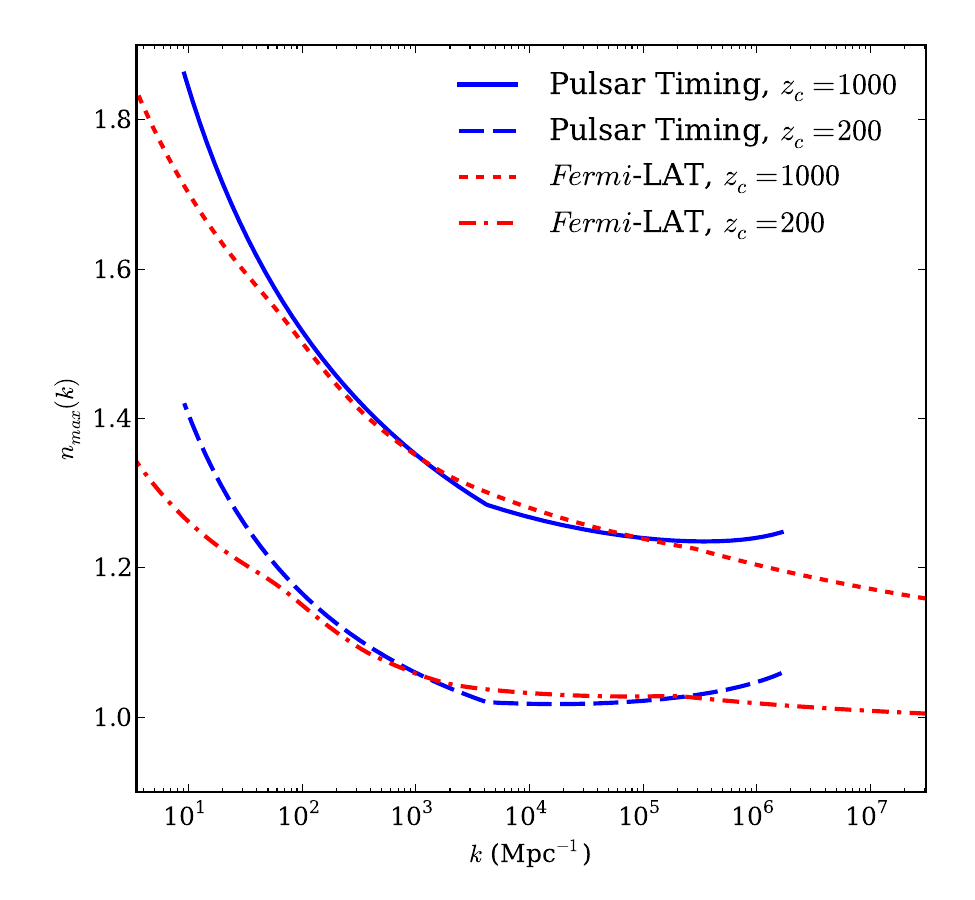}
\caption{Limits on the spectral index, $n_s$, for a scale-free primordial power spectrum, considering only constraints on $\sigma_{\chi , H}^2$ at wavenumbers smaller than $k$. These constraints are derived from 95\% CL upper limits on UCMH number density from both pulsar timing and gamma-ray searches, for two redshifts of latest collapse, $z_c$. Gamma-ray search limits assume the same dark matter model as those in Fig.\ \ref{curvlims}.}
\label{nlims}
\end{figure}

\item A stepped spectrum -- scale-free with spectral index $n_s$, with the exception of a discontinuous increase in power by $p$ at wavenumber $k_s$:
\begin{equation}
\mathcal{P}(k) \propto k^{n_s-1} \times
\begin{cases}
    1 &\textrm{ for } k < k_s\\
    p^2 &\textrm{ for } k \ge k_s
\end{cases}.
\end{equation}

In this case, we assume a constant spectral index of $n_s = 0.968$ from CMB observations by Planck, including $1\sigma$ variations allowable by their measurements. We then derive upper limits on the size of the step $p$ as a function of its position $k_s$, which are shown in Fig.\ \ref{steplims}. We find that for steps in the region $10^{0.5} \lesssim k_s \lesssim 10^6$\,Mpc$^{-1}$, the step size must be less than a factor of approximately 11 to 18 ($z_c = 1000$) or 1.5 to 2.6 ($z_c = 200$), depending upon the location of the step and the redshift of latest collapse. In contrast to these, limits from gamma-ray searches are mostly independent of the wavenumber of the step: $p_{\textrm{max}} \lesssim 10$ ($z_c = 1000$) and $p_{\textrm{max}} \lesssim 1.7$ ($z_c = 200$). 

Even if one assumes the most pessimistic case ($z_c = 1000$, $n_s = 0.974$, non-annihilating dark matter), the step size must be less than a factor of 18.4 at scales larger than $k \approx 2\times10^6 ~\textrm{Mpc}^{-1}$. Although the true upper limit is dependent upon both the redshift of latest collapse and the true value of $n_s$, our analysis has provided a strong constraint on the size of a step in primordial power at far smaller scales than previously available, independent of dark matter annihilation.
\end{enumerate}

\section{Constraints on Non-Gaussianity}
\label{NG}
Observations of the CMB suggest that the amplitudes of the primordial fluctuations follow a Gaussian distribution. However, these observations do not possess the sensitivity to rule out a distribution that is only \textit{approximately} Gaussian.  Detection of slight departures from Gaussianity would provide considerable insight into the nature of the primordial inhomogeneities. If small deviations from Gaussianity are present, the probability of larger amplitude primordial fluctuations occurring can be increased, acting to boost production of rare objects. Number counts of these objects have been shown to be sensitive to any level of deviation from a Gaussian distribution -- the rarer the object, the more sensitive it is as a probe of higher moments of non-Gaussianity (NG). Constraints on the abundances of both PBHs and UCMHs have previously been used to constrain non-Gaussianity on small-scales \citep{Bullock, Young13, Shandera}. However, these limits are either very weak (PBHs), or depend on the annihilation of dark matter (UCMHs). From constraints on the present-day number density of UCMHs in \citetalias{Hamish1}, we place limits on the level of non-Gaussianity at smaller scales than accessible via traditional methods. 

Following the method outlined in \citet{Shandera}, we express the level of non-Gaussianity in terms of a departure from the locally scale-invariant generalised Gaussian spectrum discussed in Section \ref{PPS}.  To do this, we use the model-independent dimensionless skewness, $\mathcal{M}_3$. This is a generalised form of non-Gaussianity, and may be applied to any given model. In this manner, most models that give rise to non-Gaussian interactions produce a distribution which may be expressed as an Edgeworth expansion
\begin{align}
\label{pnudnu}
P(\nu)d\nu &= \frac{d\nu}{\sqrt{2\pi}}e^{-\nu^2/2} \Bigg[ 1+\sum_{s=1}^\infty \sum_{\{k_m\}} H_{s+2r}(\nu)\nonumber \\ &\times \prod_{m=1}^s \frac{1}{k_m!} \left(\frac{\mathcal{M}_{m+2}}{(m+2)!}\right)^{k_m}\Bigg],
\end{align}
where $H_n(\nu)$ are the Hermite polynomials
\begin{equation}
H_n(\nu) = (-1)^n e^{\nu^2/2}\frac{d^n}{d\nu^n}e^{-\nu^2/2},
\end{equation}
and $\mathcal{M}_{n}$ are the dimensionless moments of the density contrast.  Here $\nu \equiv \delta_\chi/\sigma_{\chi, H}(R)$ is the `rareness' of a fluctuation in the limit of a Gaussian spectrum of perturbations.  Remembering that the mass is proportional to $R$ by Eq.\ \ref{UCMHmass}, this defines $\nu_{\rm min} \equiv \delta_\chi^{\rm min}/\sigma_{\chi, H}(R)$ as the minimum rarity required to seed the formation of a UCMH of a given mass, in the Gaussian limit.  The second sum in Eq.\ \ref{pnudnu} is over all \textit{sets} of integers $\{k_m\}$ (not members of a single set) that satisfy the equation
\begin{equation}
\label{seq} s=k_1 + 2k_2+\dotsb+sk_s.
\end{equation}
Each viable set $\{k_m\}$ implies a single value of $r$, defined as 
\begin{equation}
r=k_1 + k_2 + \dotsb + k_m.
\end{equation}

Here, higher order moments may each be expressed in terms of the third moment $\mathcal{M}_3$. We explore two types of higher-moment scaling: hierarchical and feeder, each motivated by particle physics \citep{Barnaby}. Hierarchical scaling results if the non-Gaussianity is generated by a single source, such as inflaton self-interactions or curvaton models. Otherwise, if non-Gaussian fields are coupled to the source of the curvature perturbations, then either the feeder scaling or a mixed scaling results. For the hierarchical scaling, this is expressed as
\begin{equation}
\mathcal{M}_n^h = n!~ 2^{n-3} \left(\frac{\mathcal{M}_3^h}{6}\right)^{n-2},
\label{Hscale}
\end{equation}
and for the feeder scaling, as
\begin{equation}
\mathcal{M}_n^f = (n-1)! ~2^{n-1} \left(\frac{\mathcal{M}_3^f}{8}\right)^{n/3}.
\label{Fscale}
\end{equation}

In terms of $\mathcal{M}_3$, the probability that a fluctuation of comoving radius $R$ at time of horizon entry will produce a UCMH is then
\begin{align}
\beta^{(h)}(\nu_{\rm min}) &= \textrm{erfc}\left(\frac{\nu_{\rm min}}{\sqrt{2}}\right) \nonumber \\ &+ 2\frac{e^{-{\nu_{\rm min}}^2/2}}{\sqrt{2\pi}}\sum_{s=1}^\infty \sum_{\{k_m\}_h} H_{s+2r-1} (\nu_{\rm min})\nonumber\\ &\times \prod_{m=1}^s \frac{1}{k_m!} \left(\frac{\mathcal{M}_{m+2,R}}{(m+2)!}\right)^{k_m},
\label{NGbetah}
\end{align}
for the hierarchical scaling, and
\begin{align}
\beta^{(f)}(\nu_{\rm min}) &= \textrm{erfc}\left(\frac{\nu_{\rm min}}{\sqrt{2}}\right) \nonumber \\ &+ 2\frac{e^{-{\nu_{\rm min}}^2/2}}{\sqrt{2\pi}}\sum_{s=1}^\infty \sum_{\{k_m\}_f} H_{s+1} (\nu_{\rm min}) \nonumber\\ &\times \prod_{m=1}^s \frac{1}{k_m!} \left(\frac{\mathcal{M}_{m+2,R}}{(m+2)!}\right)^{k_m},
\label{NGbetaf}
\end{align}
for the feeder scaling.  Here the integers $\{k_m\}_h$ are the non-negative solutions to Eq.\ \ref{seq}, and $\{k_m\}_f$ are non-negative integers that obey 
\begin{equation}
s+2=3k_1 + 4k_2 + \dotsb +(s+2)k_s.
\end{equation}

\begin{table}
 \caption{The reference power, $\log_{10}\mathcal{P}_\mathcal{R}^\ast$, used for each limit on non-Gaussianity from pulsar timing (Fig.\ \ref{curvlims}) and gamma-ray searches \citep{Bringmann}, for a range of scales, $k$, and redshifts of latest collapse, $z_c$.}
 \label{Pstar}
 \begin{tabular}{lccc}
  \hline
  $k$ ($\textrm{Mpc}^{-1}$) & $z_c$ & $\log_{10}\mathcal{P}_\mathcal{R}^\ast$ & $\log_{10}\mathcal{P}_\mathcal{R}^\ast$\\
  	&	& Pulsar Timing & Gamma Rays\\
  \hline
  $1\times10^1$ & 200 & -7.12 & -7.62\\
  $1\times10^4$ & 200 & -8.50 & -8.40\\
  $2\times10^6$ & 200 & -8.09 & -8.50\\
  $1\times10^1$ & 1000 & -5.41 & -5.92\\
  $1\times10^4$ & 1000 & -6.79 & -6.71\\
  $2\times10^6$ & 1000 & -6.39 & -6.87\\
  \hline
 \end{tabular}
\end{table}

\begin{figure*}
\centering
\subfloat{{\includegraphics[width=0.44\textwidth]{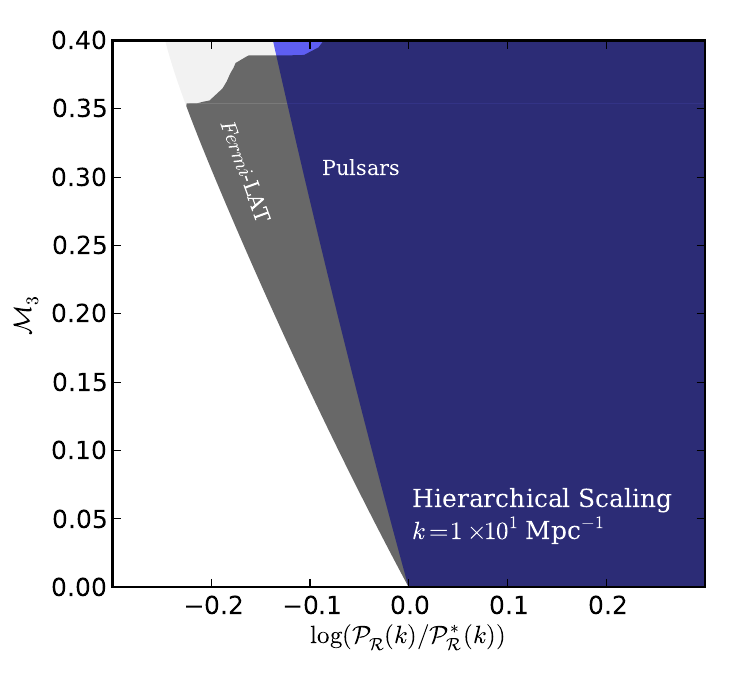}}} 
\subfloat{{\includegraphics[width=0.44\textwidth]{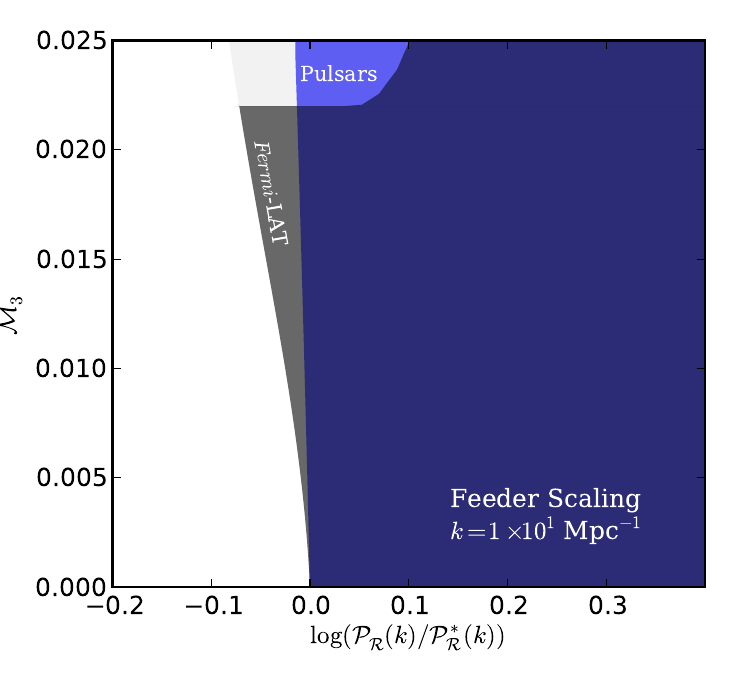}}}\\
\subfloat{{\includegraphics[width=0.44\textwidth]{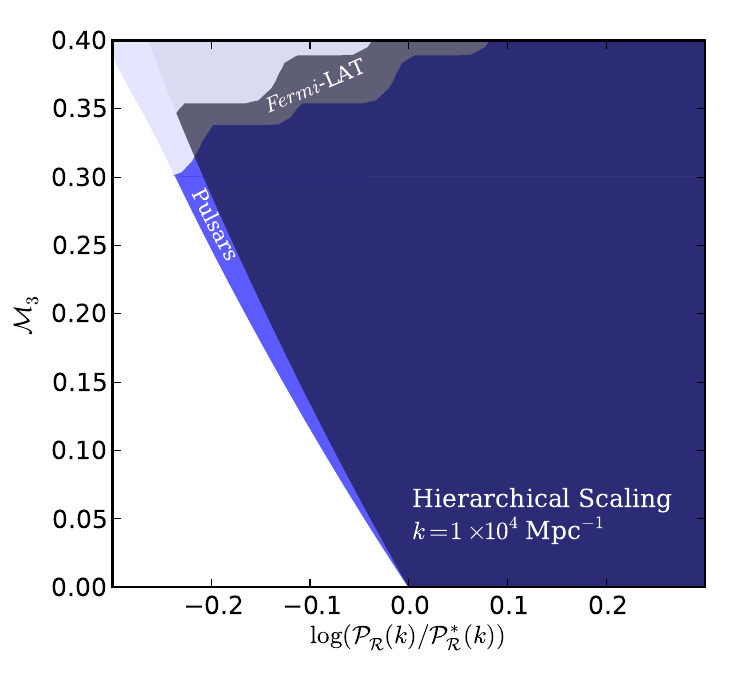}}}
\subfloat{{\includegraphics[width=0.44\textwidth]{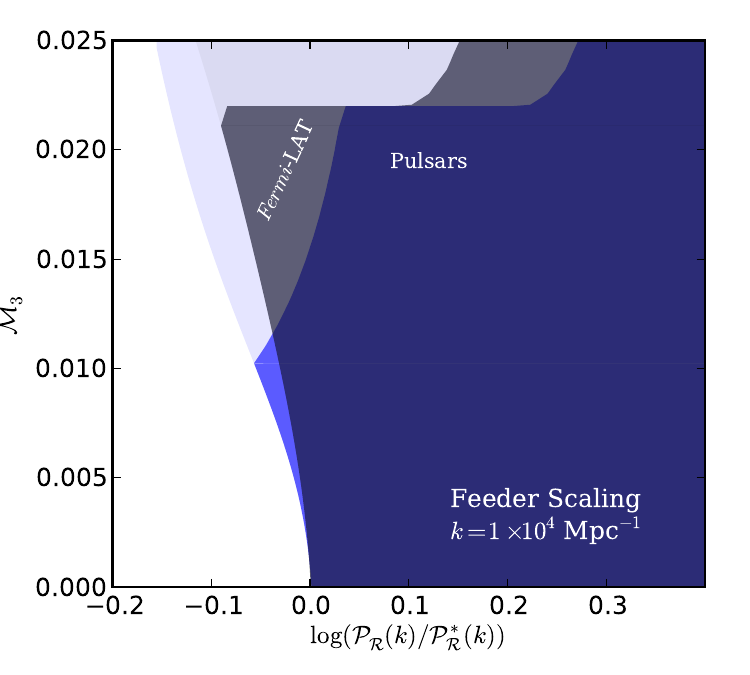}}}\\
\subfloat{{\includegraphics[width=0.44\textwidth]{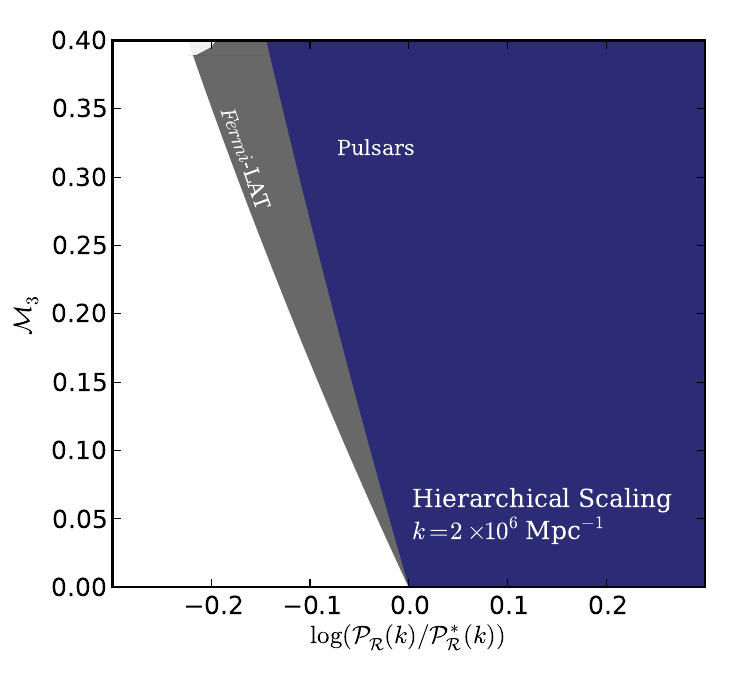}}}
\subfloat{{\includegraphics[width=0.44\textwidth]{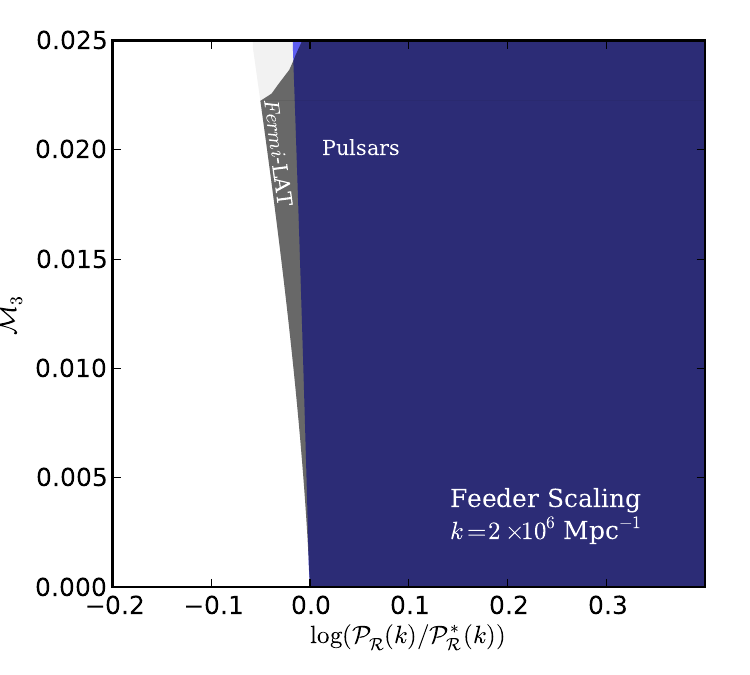}}}\\
\caption{Bounds on non-Gaussianity expressed as dimensionless skewness, $\mathcal{M}_3$, as a function of relative Gaussian power, $\mathcal{P}_\mathcal{R}(k)/\mathcal{P}_\mathcal{R}(k)^\ast$ -- given at scales near to the extrema of the limits on Gaussian power from pulsar timing in Fig.\ \ref{curvlims}. Dark shaded regions refer to those excluded at 95\% CL by either gamma-ray searches (grey) or pulsar timing (blue). Light shaded regions correspond to those that would be ruled out had the error due to the truncation of the series in eqns. \ref{NGbetah} and \ref{NGbetaf} not been accounted for. Reference power $\mathcal{P}_\mathcal{R}^\ast$ for a range of scales, $k$, and redshifts of latest allowable collapse, $z_c$, may be found in table \ref{Pstar}.}
\label{NGlims}
\end{figure*}

Again following \citet{Shandera}, we estimate the abundance of UCMHs by eqns. \ref{NGbetah} and \ref{NGbetaf}. In order to do this computationally, we must truncate the series at some finite moment of the distribution.  We discard terms with powers of $\mathcal{M}_3$ greater than 16 for the hierarchical scaling, corresponding to all terms with $s\ge17$ in Eq.\ \ref{NGbetah}, and terms with powers of $\mathcal{M}_3$ greater than 17 for the feeder scaling, corresponding to all terms with $s+2r\ge18$ in Eq.\ \ref{NGbetaf}. To compensate for this level of truncation, we exclude limits for which the estimated error can exceed 20\% (for an in-depth description of this error analysis, see section 2.3 in \citealt{Shandera}).  In order to increase the production of UCMHs, the upper tail of the distribution must be larger compared to the exactly Gaussian case.  This will occur for any positive value of $\mathcal{M}_3$.  Negative values can increase or decrease the contribution of the tail, depending upon the relative importance of odd and even $k_m$, leading to a strong dependence on the order at which the series is truncated; for this reason we show only limits on positive $\mathcal{M}_3$.

From the limits on UCMH abundance from both \citet{Bringmann} and \citet{Hamish1}, we place limits on non-Gaussianity as shown in Fig.\ \ref{NGlims} for both hierarchical and feeder scaling. We show the limits as a function of the deviation of the Gaussian power $\mathcal{P}_\mathcal{R}(k)$ from the current upper limit $\mathcal{P}^\ast_\mathcal{R}(k)$ that we found previously (i.e. Fig.\ \ref{curvlims}). We give our adopted reference powers for each wavenumber, UCMH search method, and redshift of latest collapse in Table \ref{Pstar}.

On scales of $1\times10^1$ and $2\times10^{6}$ Mpc$^{-1}$ we find that the limits from $Fermi$ are both marginally stronger and produce limits for a wider range of deviations below their Gaussian limit, compared to those from pulsar timing. Conversely, we find that on the scale of $k = 1\times 10^4 \textrm{Mpc}^{-1}$, pulsars provide a stronger limit with comparable extent to those from $Fermi$. These differences between the strength and breadth of each of these limits are minimal -- however, it must again be noted that our limits do not rely on the assumption of annihilating dark matter, and so rigorously apply to any dark matter model equally, modulo considerations of kinetic decoupling and its ability to wash out small-scale structure (see e.g. \citealt{Bringmann2009}).

\section{Constraints on Cosmic String Tension}
\label{CS}
Cosmic strings are topological defects that may have been produced in the early universe, present in many models that predict symmetry-breaking phase transitions. Their energy is confined within long, thin tubes, forming a vast network of infinite-length strings -- expected to stretch across the observable universe. When these strings cross one another (or indeed themselves), a section can detach, forming a loop. The loops oscillate, radiating gravitational waves, and so the cosmic string loops lose energy, eventually decaying away completely. These loops can gravitationally accrete matter, and thus have been shown to act as seeds for UCMH growth \citep{Berezinsky2011}. They have a complicated accretion history, dependent upon their time of formation and decay. Despite this, the number density of UCMHs expected to be formed has been predicted for given string loop radius $R$ and tension $G\mu$ \citep{Anthonisen}. Following their method, in combination with the constraints on UCMH number density from \citetalias{Hamish1}, we compute constraints on cosmic string tension as a function of loop radius.

The number density of UCMHs of a given mass produced by cosmic strings is strongly dependent upon the evolution of each string. As such, we must treat different evolution scenarios on a case by case basis, in terms of 4 critical times: time of loop formation ($x_i$), time of loop decay ($x_d$), time of latest allowed UCMH collapse ($x_c$), and the time at which UCMH accretion ceases ($x_{to}$), where time is parametrised as
\begin{equation}
x(t) \equiv \frac{a(t)}{a(t_{eq})} = \frac{z_{eq} + 1}{z(t)+1}.
\end{equation}

With this parametrisation, the redshift of matter-radiation equality, $z_{eq}$, corresponds to $x=1$. In what follows, we take the assumption that $x_{to}$ refers to the time after which structure formation has progressed sufficiently to allow the majority of UCMHs to be within bound structures, preventing further accretion from the smooth cosmological background (as discussed in Section \ref{PPS}): $z_{to} \approx 10$, $x_{to} \approx 284$. Similar to the previous sections, we examine the case of redshifts of latest collapse of both $z_c = 1000$ and $z_c=200$, corresponding to $x_c = 3.12$ and $x_c = 15.54$, respectively.

We follow a one-scale loop model \citep{Vilenkin1981,Kibble1985}, which describes loops of a given radius as being produced together at the same time. The other critical times $x_i$ and $x_d$ are then dependent upon the properties of cosmic strings formed at each epoch, as:
\begin{align}
x_i &= \left(\frac{\beta}{\alpha}\frac{R}{t_{eq}}\right)^a,\\
x_d &= \left(\frac{\beta}{\gamma G\mu}\frac{R}{t_{eq}}\right)^a,
\end{align}
where $a=1/2$ when $x < 1$, and $a=2/3$ when $x>1$, and $\alpha=0.05$, $\beta=2\pi$, and $\gamma=10\pi$ are constants determined from simulations \citep{Vachaspati1985,BP2011}.

The fractional density of UCMHs of a given mass is related to the properties of cosmic strings formed at a particular time by
\begin{equation}
\frac{df_{MW}}{dM_h^0} = C \rho_{DM}^{-1} n(R,t) R,
\end{equation}
where $\rho_{DM}$ is the present day density of dark matter in the universe, $C$ is a constant dependent upon the time of formation and decay of the loop,
\begin{alignat}{3}
&C(x_i < 1, x_d <1) &&= \frac{2+3x_d}{3+3x_d},\\
&C(x_i < 1, x_d >1) &&= 1,\\
&C(x_i > 1, x_d < x_c) &&= \frac{6x_{to}(x_i^{-1}-x_d^{-1})-9}{2x_{to}(x_i^{-1}-x_d^{-1})-9},\\
&C(x_i < 1, x_d <1) &&= \frac{6x_{to}-15x_i}{2x_{to}-15x_i},
\end{alignat}
and $n(R,t)$ is the number density (ignoring decay) of loops of a given radius at some time $t$. During radiation domination this is
\begin{equation}
n(R,t) = N \alpha^2 \beta^{-2} t^{-2} R^{-2},
\end{equation}
and during matter domination
\begin{equation}
n(R,t) = N \alpha^{5/2} \beta^{-5/2} t_{eq}^{1/2} t^{-2} R^{-5/2},
\end{equation}
where $N$ is another constant to be estimated by simulation.  We take $N=40$, as seen by \citet{BP2011}.

While these calculations assume that each loop is stationary, cosmic string loops are expected to be formed possessing relativistic velocities \citep{BP2011}. This non-zero velocity will decrease the efficiency of UCMH accretion, as the infall of matter no longer occurs in a spherically symmetric manner. This can be simplistically accounted for by assuming that, if a loop were to travel further than some distance $KR$ before decaying, a UCMH will not be formed. The differential UCMH fraction will then be suppressed by a factor of $\mathcal{S}$, resulting in:
\begin{equation}
\frac{df_{MW}}{dM_h^0} = \mathcal{S}\frac{16\pi G C N \alpha^2}{3 R \beta^2 f_\chi \kappa} X^{1/2},
\label{dfdm}
\end{equation}
where $X=1$ for loops formed after matter-radiation equality ($x_i > 1$) or $X =\alpha t_{eq}/(\beta R)$ for those formed before ($x_i<1$), $G$ is the gravitational constant, and
\begin{equation}
\kappa \equiv H_{eq}^2t_{eq}^2 = \frac{16\pi G \rho_{DM}(t_{eq})t_{eq}^2}{3f_\chi}.
\end{equation}

For a particular initial velocity $v_i$ of a cosmic string loop, the suppression factor will be:
\begin{equation}
\mathcal{S} = \frac{2^{1/2} v_i^3}{3\pi^{1/2}\langle v^2 \rangle^{3/2}}.
\end{equation}
where we again follow \citet{Anthonisen} by taking $\langle v^2 \rangle^{1/2} = 0.3$, which assumes that the loop velocity distribution follows that of the long strings. This suppression factor may be expressed in terms of $K$ by:
\begin{equation}
v_i < K \frac{(\alpha \gamma G \mu)^{1/2}}{\beta \ln\left( \frac{\alpha}{\gamma G \mu}\right)}.
\end{equation}

\begin{figure}
\centering
\includegraphics[width=\columnwidth]{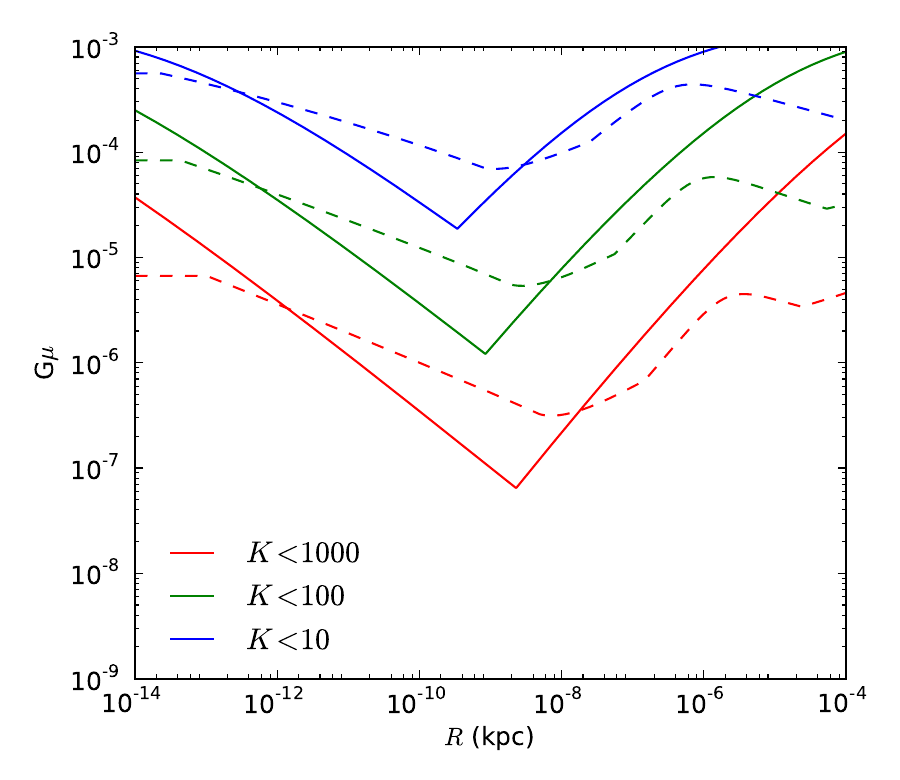}
\caption{Constaints on cosmic string tension $G\mu$ as a function of loop radius, for a range of different velocity suppression factors $K$. We display limits derived from those on UCMH number density from both pulsar timing (solid lines) and gamma-ray searches (dashed lines). Limits from gamma-ray searches again assume dark matter mass $m_\chi = 1$ TeV, and 100\% annihilation into $b\bar{b}$ pairs with cross-section $\langle \sigma v \rangle = 3\times 10^{-26} \textrm{cm}^3\textrm{.s}^{-1}$.}
\label{cslims}
\end{figure}

By combining Eq.\ \ref{dfdm} with the constraints on the UCMH abundance from gamma-rays and pulsar timing, we are able to place limits on the cosmic string tension for a given loop radius and suppression factor (parametrised in terms of $K$). We plot these constraints in Fig.\ \ref{cslims} for several values of the constant $K$, adopting a redshift of latest collapse of $z_c=1000$. As this derivation depends on the cosmic string scaling solution, it is important to note that the actual resulting limit on $G\mu$ corresponds to the strongest limit at any $R$. As such, not only do the constraints due to pulsar timing remove the assumption that dark matter must annihilate, but also strengthen the overall best constraint on string tension from $G\mu \leq 3.14 \times 10^{-7}$ to $G\mu \leq 6.49 \times 10^{-8}$ for loops that are able to travel 1000 times their own radius and still form a UCMH. 

The shape of these limits may be understood by comparison with the constraints on UCMH number density. For subhalo masses greater than approximately $10^3 M_\odot$, these limits are saturated by the probability that there are insufficient UCMHs within the Milky Way to provide a reliable signal. The strongest limit at this point is mapped to a constraint on the mass of the loop, which in turn is proportional to its tension and radius as $M = \mu\beta R$. This mapping translates the peaked shape of constraints on UCMH abundance directly to those found here.

\section{Conclusion}
\label{conc}

The large-scale structure of the universe is thought to have been seeded by small fluctuations in the early universe. Although these structures were formed from overdensities of order $\delta \sim 10^{-5}$, larger amplitude fluctuations are thought to be able to produce rare structures such as primordial black holes ($\delta \gtrsim 0.3$) and ultracompact minihalos ($0.3 \gtrsim \delta \gtrsim 10^{-3}$).

The abundance of these rare objects has been used to constrain a variety of processes that would boost their production beyond that expected from the standard Harrison-Zel'dovich (scale-free) model. To date, three such processes have been constrained: an increase in primordial power at small scales, deviations of the distribution of primordial fluctuations from a Gaussian, and the presence of cosmic strings in the early universe. By consideration of each process in turn, it is possible to link their properties to the present day number density of each rare object.

While previous studies have given constraints on both UCMH and PBH number densities, these are either very weak, or strongly dependent on assumptions about the specific particle nature of dark matter. By considering new upper limits on the number density of UCMHs \citepalias{Hamish1}, we provide updated constraints on the properties of each of these processes.  Although the limits are strongly dependent upon the assumed redshift of latest formation of a UCMH, $z_c$, even a very conservative value of $z_c=1000$ results in some of the strongest limits to date.

Here we have calculated the contribution of increased primordial power on small scales for 3 different power spectrum models. For a `generalised' power spectrum, we found $\log_{10}\mathcal{P}_\mathcal{R} \lesssim -6.5$ ($z_c = 1000$) and $\log_{10}\mathcal{P}_\mathcal{R} \lesssim -8.5$ ($z_c = 200$), in the range $10^1 \lesssim k \lesssim 10^7$.  This is comparable to the limits from non-detection of dark matter sources by $Fermi$-LAT. We additionally find limits of $n_s \leq 1.24$ ($z_c = 1000$) and $n_s \leq 1.02$ ($z_c = 200$) on the spectral index of a scale-free power spectrum. For a stepped spectrum, the non-observation of UCMHs limits the step size to a factor of approximately 11--18 ($z_c = 1000$) and 1.5--2.6 ($z_c = 200$).

We also provide limits on the dimensionless skewness $\mathcal{M}_3$, dependent upon both the scale of the fluctuation, $k$, and redshift of latest collapse, $z_c$. Assuming two different models of scaling with higher moments, we find limits that are independent of dark matter annihilation, and (depending on the scale) are able to be applied to lower primordial power, and to wider variations. Depending on the nature of the non-Gaussianity, these limits can be to be easily mapped to the more model-dependent quantity $f_{\rm NL}$, for comparison. For example, if we take simple non-linear coupling, $\mathcal{R}(x) = \mathcal{R}_{\rm G}(x) + \frac{3}{5}f_{\rm NL}\left[\mathcal{R}_{\rm G}(x)^2 - \left<\mathcal{R}_{\rm G}(x)^2\right>\right]$, the two may be related by $f_{\rm NL} \approx \mathcal{M}_3/\mathcal{P}_\mathcal{R}^{1/2}$. This results in a constraint of $f_{\rm NL} \lesssim$ 
$\mathcal{O}(10^2)$ to $\mathcal{O}(10^3)$, depending upon the exact shape of the primordial power spectrum.

Finally, we constrain cosmic string tension as $G\mu \leq 6.49 \times 10^{-8}$, under the assumption that a loop can move up to $K = 1000$ times its own radius and still form a UCMH. Although this constraint is stronger than that from CMB observations ($G\mu \leq 1.7\times 10^{-7}$ from \citealt{Dvorkin}), this assumed value of $K$ is probably overly optimistic. As $K$ decreases, the limit grows significantly weaker. While far stronger limits of $G\mu \leq 2.8\times 10^{-9}$ were obtained by \citet{Blanco-Pillado2014}, their result relies on the proper understanding of emission of gravitational waves from cosmic string cusps. As this is a poorly understood process, these constraints must be treated with appropriate caution.

We have shown that these limits depend heavily on the latest redshift at which UCMHs are assumed to be able to form. We have displayed constraints for both the conservative value of $z_c =1000$ and the more optimistic $z_c = 200$. It is important to note that there is, as yet, no strong evidence in favour of either value. Should further research be undertaken to investigate the physical value of $z_c$, the limits here could potentially be improved substantially, leading to the exclusion of multiple cosmological models.

\section*{Acknowledgements}
HAC would like to acknowledge the Australian Postgraduate Awards (APA), through which this work was financially supported. GFL gratefully acknowledges the Australian Research Council (ARC) for support through DP130100117. PS is supported by STFC through the Ernest Rutherford Fellowships scheme.
\clearpage




\bibliographystyle{mnras}
\bibliography{UCMHPaper2.bib} 
\bsp	
\label{lastpage}

\includepdf[pages={1,2,3,4}]{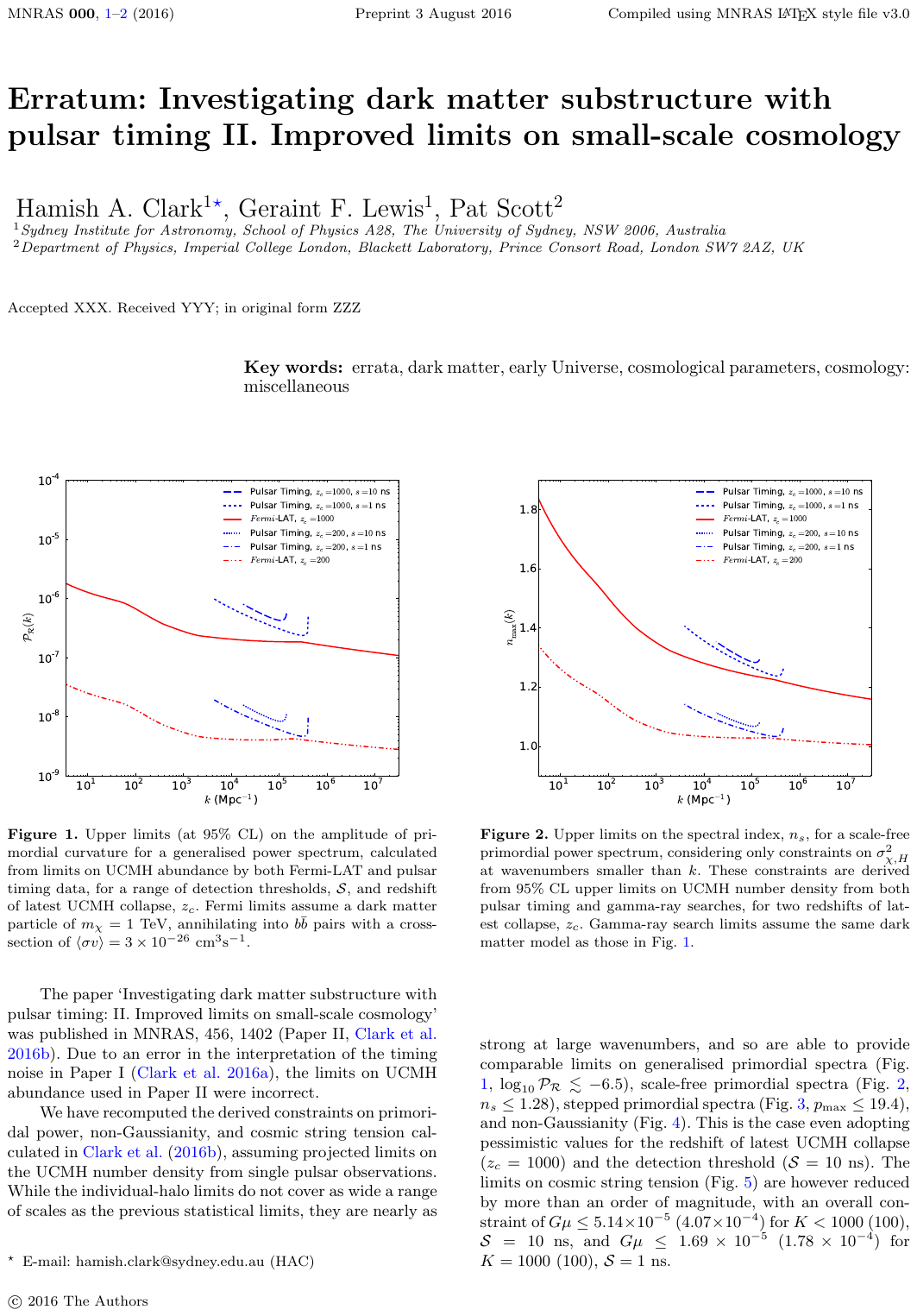}
\end{document}